\documentclass[12pt]{article}

\newcommand{\PSbox}[3]{\mbox{\rule{0in}{#3}\includegraphics{#1}\hspace{#2}}}

0

\begin{document}
\newcommand {\be}{\begin{equation}}
\newcommand {\ee}{\end{equation}}
\newcommand {\bea}{\begin{array}}
\newcommand {\cl}{\centerline}
\newcommand {\eea}{\end{array}}
\def\simlt{\stackrel{<}{{}_\sim}}
\def\simgt{\stackrel{>}{{}_\sim}}
\def\IP{\relax{\rm I\kern-.18em P}}
\def\CI{case {\bf (I) }}
\def\CII{case {\bf (II) }}
\def\sugra{supergravity }

\renewcommand {\theequation}{\thesection.\arabic{equation}}
\renewcommand {\thefootnote}{\fnsymbol{footnote}}
\newcommand {\newsection}{\setcounter{equation}{0}\section}
\csname @addtoreset\endcsname{equation}{section}

\def\ba{\begin{eqnarray}}
\def\ea{\end{eqnarray} }
\def\o{\over}  
\def \a {\alpha }
\def\th{\theta }
\def\s{\sigma}
\def\t{\tau }
\def\p{\partial }
\baselineskip 0.65 cm
\begin{flushright}
IC/2000/73 \\
hep-th/0006202
\vskip 14mm   
\end{flushright}

\begin{center}
{\LARGE{ On Noncommutative  Open String Theories }}
\vskip 14mm

{\bf {\large{J. G. Russo$^{1,2}$  and M. M. Sheikh-Jabbari$^2$}}}  
% $^{\dag}$

\vspace{12 mm}

$^1${\it Departamento de F\'\i sica, Universidad de Buenos Aires\\
Ciudad Universitaria, 1428 Buenos Aires, and Conicet}\\
{\tt russo@df.uba.ar}\\

$^2${\it The Abdus Salam International Center for Theoretical Physics\\
 Strada Costiera, 11. 34014, Trieste, Italy}\\
{\tt  jabbari@ictp.trieste.it }\\

\end{center}

\vskip 8mm
\begin{center}
{\bf Abstract}
\end{center}

\vspace{2 mm}

We investigate new compactifications of  OM theory 
giving rise to
a 3+1 dimensional  open string theory with noncommutative $x^0$-$x^1$ and
$x^2$-$x^3$ coordinates.
The theory can  be directly obtained 
by starting with a D3 brane with parallel
(near critical) electric and magnetic field components, in the presence of
a RR scalar field.
The magnetic parameter permits to interpolate continuously between
the  $x^0$-$x^1$ noncommutative open string theory and the
$x^2$-$x^3$ spatial noncommutative  $U(N)$ super Yang-Mills theory.
We discuss  $SL(2,Z)$ transformations of this theory.
Using the supergravity description of the large $N$ limit,
we also compute  corrections to the quark-antiquark Coulomb potential
arising in the NCOS theory.

% they lead to
% stronger attraction forces at short distances.

\vspace{3 cm}

%{\footnotesize \dag}\hspace{1 mm}{\footnotesize{\ttfamily e-mail address:
% jabbari@ictp.trieste.it and  ....}}

\newpage

%%%%%%%%%%%%%%%%%%%%%%%%%%%%%%%%%%%%
\newsection{Introduction }
%%%%%%%%%%%%%%%%%%%%%%%%%%%%%%%%%%%%

Recently there has been an increasing interest in
noncommutative geometries \cite{{CDS},{SW}}, where coordinates 
satisfy
\be
[x^{\mu},x^{\nu}]=i\theta^{\mu\nu} \ .
\ee
It has been shown that theories on such spaces naturally arise in string
theory as the worldvolume theory of $Dp$-branes in a $B_{\mu\nu}$-field
background \cite{{SW},{AAS},{Ho}}. 
When $\theta^{0\mu}=0$, then 
 the low-energy effective theory on such branes is a noncommutative
Super Yang-Mills (NCSYM) theory  with 16
supercharges \cite{{SW},{Sh}}. 
In particular, starting with  $N$ coincident D3-branes along $0123$
directions with a non-zero $B_{23}$ field, one 
obtains a ${\cal N}=4,\ D=4$ noncommutative $U(N)$ theory. 
%Loop calculations show
%that these theories have a zero beta function, 
%just like their commutative counterparts \cite{}.
%In addition to the perturbative loop expansions, 
The large $N$ limit of the 
NCSYM theories have been studied
through the generalization of gravity/gauge theory correspondence
 to the cases with $B$-fields \cite{HI,MR,AOS}.
%However, in this way limit of the theory with a large, 
% but finite, effective 't Hooft coupling 
% ($g^2_{eff}=Ng^2_{YM}$) have been studied. 

The case $\theta^{0\mu}\neq 0$ was recently studied in \cite{SST,GMMS},
and it is relevant to the strong coupling limit of 
NCSYM theories.
% with fixed $N$. 
This point was investigated in \cite{GMMS},
 using the
strong-weak coupling duality symmetry of type IIB string theory,
which transforms  the $B_{23}$ (magnetic) background field
into a $B_{01}$ (electric) background. 
As a result, the strongly coupled theory can be described in terms of a weakly coupled open string
theory where the coordinate $x^1$ and the time coordinate $x^0$ are
noncommutative.
The emergence of an open string theory --rather than a field theory--
is not unexpected, since there are arguments  indicating that
 a field theory in a noncommutative spacetime ($\theta_{0i}\neq 0$) cannot be unitary \cite{Gom}.
The resulting open string theory, which can be directly obtained by starting with a system of 
D3 branes in the presence of a near critical $B_{01}$ field,
has indeed a string scale $\a '$ which is of the same order of the
noncommutative scale $\theta $,
so string effects cannot be disentangled from noncommutative effects.
% the $\theta_{0i}\to 2\pi\alpha'$, these terms become equally important. 
For a near critical electric field, one can show that closed string states all become very
massive and decouple from the theory. 
One ends up with a consistent  theory of open strings which does not require closed strings,
noncommutative open string theory (NCOS) (for other recent works on
spacetime noncommutativity see \cite{{Ganor},{Barbon},{Harm},{1+1}}).

A similar construction can be made for other Dp-branes, and also by starting with the M5 brane
in a near critical three-form field strength. It is believed that the later
leads to a theory of  open membranes  (OM theory)
\cite{OM,BBSS},
which is decoupled from eleven dimensional gravity.
One obtains in this way a web of open string/membrane theories connected by dualities,
much like the web of dualities connecting string and M-theory
compactifications \cite{OM}.
These theories can thus be used to explore string phenomena without the complications of gravity.

The aim of this work is to explore another piece of this web by
considering a D3 brane system in the presence of both  $B_{01}$ and 
$B_{23}$ field components, 
which is equivalent to a configuration of parallel electric
and magnetic vector fields on the brane
(a discussion on electric and magnetic fields
suggesting  a new type of dualities involving Lorentz boosts
 can be found in \cite{chen}).
We shall show that there is a limit (corresponding to a critical value of the $B_{01}$ component) 
in which  closed string modes are
decoupled, leading to a family of noncommutative open string theories in
$x^0$-$x^1$ and $x^2$-$x^3$ directions. 
%The extra parameters permit to
%interpolate between
%the pure electric case and its strong coupling dual magnetic 
% NCSYM theory.
When the noncommutative length scale in the $x^2$-$x^3$ direction is
greater than the string scale $\a '$,
for energies much lower than $\a '$, one obtains a   
NCSYM theory in $x^2$-$x^3$
directions, generically with a non-zero $\Theta $ term (i.e. of the form 
$\Theta \int \tilde F_{\mu\nu}F^{\mu\nu}$).
We study these theories under the
$SL(2,Z)$ symmetry of type IIB superstring theory.
For  irrational $\Theta$, a generic
$SL(2,Z)$ transformation  maps our NCOS theory to another NCOS theory with
different parameters
(in particular, different noncommutative scales).
For  rational $\Theta $, there is a (one parameter) 
family of $SL(2,Z)$ rotations which map the NCOS theory to a NCSYM theory.

The paper is organized as follows. 
In section 2, we consider a D3-brane in presence of both electric
and
magnetic backgrounds, and determine the spectrum of open and closed 
strings.
We show that there is a critical electric field limit for which
the closed string modes decouple. 
Using  the type IIB supergravity solution corresponding to
the
D3-brane configuration with $B_{01}$ and $B_{23}$, in section 3
we  study the  transformation properties of these general NCOS theories
induced by   the $SL(2,Z)$ symmetry of the type IIB
theory. 
Section 4 contains another derivation of  these NCOS theories 
by compactifying OM theory along two arbitrary directions
of a 2-plane.
% The magnetic  and scalar RR field parameters are related to 
% moduli.
In section 5, we use the supergravity solutions which are
the holographic dual of the NCOS theories to obtain 
a quark-antiquark potential in the NCOS theory at large $N$.

%%%%%%%%%%%%%%%%%%%%%%%%%%%%%%%
\newsection{Open strings in the presence of $B_{01}$ and $B_{23}$}
%%%%%%%%%%%%%%%%%%%%%%%%%%%%%%%

\def\lm{\left( \matrix{ }
\def\rm{ }\right) }

Let us consider a D3 brane in the presence of a constant $B$-field
with components $B_{01}$ and $B_{23}$.
Such constant $B$-field is equivalent to  constant electric
and magnetic vector fields on the brane  pointing in the direction $x^1$.
Open strings in constant gauge fields were
originally discussed in \cite{abou}.
Let $x^\mu $, with $\mu, \nu =0,1,2,3$, label the directions that are parallel
to the brane, and $x^a $, $a,b=4,...,9$, the transverse directions.
It is convenient to choose coordinates so that the closed string metric
has the form
$g_{\mu\nu}={\rm diag}( -\zeta,\zeta , \eta, \eta)$, $g_{ab}=\delta_{ab}$,
where $\zeta $ and $\eta $ are constant parameters which will be fixed
later.
%\be
%g_{\mu\nu}= \left(\matrix { -\zeta  & 0 & 0 & 0 \cr 
%                 0  & \zeta & 0 & 0 \cr
%                 0  & 0    & \eta & 0 \cr
%                 0 &  0    & 0    & \eta }\right)
%\ee
The boundary conditions for the open string coordinates are given by
\ba
& \partial_{\sigma}X^0+E \partial_{\tau}X^1 \ \bigg|_{\s=0,\pi }=0 \ ,\ \
\ \ 
\partial_{\sigma}X^1+E \partial_{\tau}X^0\ \bigg|_{\s=0,\pi }=0 \ , \\
& \partial_{\sigma}X^2+B \partial_{\tau}X^3\ \bigg|_{\s=0,\pi }=0 \ ,\ \ \
\ 
\partial_{\sigma}X^3-B \partial_{\tau}X^2\ \bigg|_{\s=0,\pi }=0 \ ,
\ea
$$
E\equiv  B^0_{\ 1}=- \zeta^{-1} B_{01}\ ,\ \ \ \ \
B\equiv B^2_{\ 3}=\eta ^{-1} B_{23}\ .
$$
The transverse open string coordinates are free string coordinates obeying
Dirichlet boundary conditions:
$$
X^a\ \bigg|_{\s=0,\pi }=x^a\ ,\ \ \ \ a=4,...,9\ ,
$$
where $x^a$ denote the position in transverse space of $N$ D3 branes.
The solutions to the above boundary conditions are
$$
X^0 =x^0+2\a ' (p^0\t - E p^1 \s ) +\sqrt{2\a '} \sum_{n\neq 0} {1\over
n}e^{-in\t }\big[ i
a^0_n
\cos(n\s
)- E a^1_n \sin(n \s) \big]\ , 
$$
$$
X^1 =x^1+2\a ' (p^1\t - E p^0\s ) +\sqrt{2\a '} \sum_{n\neq 0} {1\over
n}e^{-in\t }\big[ i
a^1_n
\cos(n\s
)- E a^0_n \sin(n \s)\big] \ , 
$$
\be
X^2 =x^2+2\a ' (p^2\t -B p^3\s )+\sqrt{2\a '} \sum_{n\neq 0} {1\over
n}e^{-in\t }\big[ ia^2_n
\cos(n\s
)- B a^3_n \sin(n \s)\big] \ , 
\label{OS}
\ee
$$
X^3 =x^3+2\a ' (p^3\t + B p^2\s )+ \sqrt{2\a '} \sum_{n\neq 0} {1\over
n}e^{-in\t }\big[ ia^3_n
\cos(n\s
)+ B a^2_n \sin(n \s)\big] \ , 
$$
$$
X^a=x^a+ \sqrt{2\a '}\sum_{n\neq 0} {\alpha _n^a \over n} e^{-i n\tau }
\sin (n\s )\ .
$$

%%%%%%%%%%%%%%%%%%%%%%
   
\def\th{\theta }
\def\td{\tilde \theta }

It is convenient to introduce the ``open string parameters" $G_{\mu\nu},\
\theta_{\mu\nu},\ G_s$   \cite{SW}  as follows:
\be\label{ggg}
G_{\mu\nu}= g_{\mu\nu} - (B g^{-1} B)_{\mu\nu}=
\left(\matrix{ -\zeta (1-E^2)  & 0 & 0 & 0 \cr
0  & \zeta (1-E^2) & 0 & 0 \cr
0  & 0    & \eta (1+B^2) & 0 \cr
0 &  0  &   0  & \eta (1+B^2)  }\right)\ ,
\ee 
\be\label{thth}
\theta^ {\mu\nu} = 2\pi \a' \left( {1\over g+ B }\right)^{\mu\nu}_A
=\left(\matrix{ 0& \th  & 0 & 0  \cr
  -\th  & 0 & 0 & 0 \cr
   0  & 0  & 0 & \td  \cr   
  0 &  0  & -\td & 0 }\right)\ ,
\ee
\be\label{Ocoup}
G_s=g_s \left({\det G_{\mu\nu}\over \det (g_{\mu\nu}+B_{\mu\nu })}
\right)^{1\over 2} = 
g_s \sqrt{(1-E^2)(1+B^2) }\ ,
\ee
where
\be
\th = 2\pi\alpha '{E\over \zeta (1-E^2) }\ ,\ \ \ \ \td =
2\pi \a '{B\over \eta (1+B^2)}\ .
\ee
The canonical commutation relations for the string coordinates then imply the
following commutation
relations for the mode operators:
$$
[a_n^\mu ,a_m^\nu ]=n \delta_{n+m} G^{\mu\nu}\ ,
$$
$$
[x^\mu,x^\nu ]= i\theta ^{\mu\nu}\ ,\ \ \ \ \ [x^\mu,p^\nu ]=i G^{\mu\nu}\
.
$$
One can also check that the ends of the string do not commute, i.e.
$$
[X^\mu (\tau, 0),X^\nu (\tau, 0)]=i \theta^{\mu\nu}\ ,\ \ \ \ \ 
[X^\mu (\tau, \pi),X^\nu (\tau, \pi)]=-i \theta^{\mu\nu}\ .
$$
Using (\ref{OS}), one can obtain the Hamiltonian operator: 
$$
L_0={1\over 4}\a ' G^{\mu\nu } p_\mu p_\nu +2 \sum_{n=1}^\infty G_{\mu\nu}
a_{-n}^\mu a_n^\nu +2 \sum _{n=1}^\infty \a_{-n}^a\a_{n}^a -1
\ .
$$

Now we set 
\be\label{zeta}
\zeta =(1-E^2)^{-1}\ , \ \ \ \  \eta =(1+B^2)^{-1}\ ,
\ee
so as to have 
\be
G_{\mu\nu}=\eta_{\mu\nu} \ , \ \ \  \ \theta=2\pi\alpha' E\ ,\ \ \ \
\tilde\theta=2\pi\alpha' B\ .
\ee
The open string mass spectrum then becomes manifestly the same as the
free string mass spectrum:
$$
\a' M^2=\a' (p_0^2-p_1^2-p_2^2-p_3^2)= N-1\ .
$$
The addition of fermions is as in the usual free
open superstring theory (with the appropriate change in the
normal ordering constant).
The closed string mass spectrum is then given by
$$
\a' (1-E^2)  (p_0^2 -p_1^2 )- \a' (1+B^2) (p_2^2+p_3^2)= 2N+2\bar N-4\ .
$$
We see that as $E\to 1 $ the energy of the closed string states
goes to infinity, while the energy of open string excitations
remains finite.
In the limit $E\to 1 $ with fixed $B$, $\a '$ and $G_s$,
the closed string states are thus decoupled from the theory.
This is just as in the zero magnetic field case discussed in \cite{SST,GMMS}.
Note that this limit requires $g_s\to \infty $ (see eq.~(\ref{Ocoup})~).

The resulting open string theory obtained in this
limit contains the parameters
$G_s $, 
\be
\theta= 2\pi \a'   \ ,\ \ \ \  \td =2\pi \a' B=\th B\ .
\label{uuuu}
\ee
In the next section it will be shown  that   one can also introduce 
another parameter $\chi$ associated with the RR scalar of type IIB theory.
So, altogether our NCOS theory is defined by four parameters, $\alpha',\tilde\theta,
G_s$ and $\chi $. 
The parameter $\alpha'= \theta /2\pi $ is the string scale and also characterizes the
noncommutative scale in the $x^0$-$x^1$ directions, $\tilde \theta $ represents the noncommutative scale in the
$x^2$-$x^3$ directions,  $G_s$ is the open string coupling, while $\chi $ is not relevant  in
the perturbative expansion.

General scattering amplitudes on the disc for this ``noncommutative"
open string theory will have the form
\be
\langle V(p^{1})...V(p^{N}) \rangle _{\theta,\td }
=\exp\big[ -{i\over 2} \sum_{n>m}
p^n\wedge p^m \epsilon(\tau_n-\tau_m)\big] \ 
\langle V(p^{1})...V(p^{N}) \rangle _{\rm free\ string}\ ,
\ee
where
$$
p^n\wedge p^m= (p_0^{n} p_1^{m}-p_1^{n} p_0^{m})
\th + (p_2^{n} p_3^{m}-p_3^{n} p_2^{m}) \td \ . 
$$
In the case $B=0$, one recovers the open string theory of
\cite{SST,GMMS},
obtained from open strings in a purely electric background.

A special limit of this NCOS theory is given by 
 $B\to \infty,\ \alpha'\to 0$, with
\be
G_s={\rm fixed}\ ,\ \ \ \ \alpha'B={\rm fixed}\ .
\ee
% Since the closed string coupling is very small the gravity interactions with the
% bulk fields can be effectively neglected and hence 
Because $\a'\to 0$,  massive open string excitations also decouple in this limit,
and one is left with
the Super Yang-Mills field theory in  noncommutative $x^2$-$x^3$ space.
% In general, there will also be  a $\Theta$ term.
Thus the present family of NCOS theories  interpolates  between 
the purely electric NCOS theory of \cite{SST,GMMS} and the NCSYM.

In general, the theory contains two energy scales, given by
$\theta $ and $\tilde \theta $, or $\a' $ and $B\a' $.
At distances  $L$ much larger than $\sqrt{\theta }$, $\sqrt{\td }$,
the theory reduces to ordinary SYM theory. For $B\gg 1$, there is a
regime $\theta \ll L^2 <\td $ in which the theory is  SYM field
theory on the noncommutative space $x^2$-$x^3$; string effects
can be ignored, but noncommutativity effects in $x^2$-$x^3$ directions are
important. If $B$ is of order 1 or lower, then string effects appear
at the same time as noncommutative effects.

%%%%%%%%%%%%%%%%%%%%%%%%%%%%%%%%%%%%%%%%%%%%%%%%%%%%%
\newsection{SL(2,Z) transformations on NCOS theory}
%%%%%%%%%%%%%%%%%%%%%%%%%%%%%%%%%%%%%%%%%%%%%%%%%%%%%%%%%

In this section we study the behavior of the open string theories of
the previous section
under type IIB $SL(2,Z)$ symmetry. 
% To obtain the $SL(2,Z)$ rotated open
% string parameters and
% also to find the fourth moduli of our theory, RR scalar ($\chi$), 
We start with the
corresponding type IIB supergravity solution.
The Lorentzian supergravity solution representing a D3 brane in the
presence of
$B_{01}$ and $B_{23}$ fields is given  in \cite{MR}
(see also \cite{ddff}).
In the string frame, it is 
\be
ds_{str}^2 =  f^{-1/2} \bigg[ h'(- d{x_0}^2 +
 d{x_1}^2) + h (d{x_2}^2 +d{x_3}^2)\bigg] 
+f^{1/2} \bigg[ dr^2 + r^2 d\Omega_5^2 \bigg]\ ,
\label{susu}
 \ee
$$
f  = 1 + { {\alpha'} ^2 R^4 \over r^4 } ~, ~~~~~
{h}^{-1} = \sin^2\alpha f^{-1}+ {\cos^2\alpha } \ , ~~~~~ 
{h'}^{-1} = -\sinh^2\beta f^{-1}+ {\cosh^2\beta } \ ,
$$
$$
B_{01} = - \tanh\beta f^{-1} h'  \ ,\ \ \ \ \ \
B_{23} =  \tan\alpha f^{-1} h  \ ,
$$
$$  
e^{2\phi} = {g_s}^2 h h'\ ,\ \ \ \ \
\chi = {1 \over g_s } \sinh\beta\sin\alpha f^{-1}\  +\chi_0, 
$$
$$
A_{01} = ({1 \over g_s} \sin\alpha \cosh\beta +\chi_0 \tanh\beta)
h'f^{-1} \ , \ \ \
A_{23} = ({1 \over g_s} \sinh\beta \cos\alpha -\chi_0 \tan\alpha) hf^{-1}
,
$$
$$
F_{0123u}  = {1 \over g_s}\cos\alpha \cosh\beta \  h h' \partial_r f^{-1}
\ .
$$ 
We have generalized the solution of \cite{MR} by adding
an extra parameter $\chi_0$ obtained by the $SL(2,R)$ transformation
$\chi\to\chi+ \chi_0$.

In the $r\to\infty $ region, the metric  asymptotically approaches the
Minkowski metric, and the
asymptotic  values for the different  fields
are as follows
\be
(B^{\infty})^{0}_{\ 1} = \tanh\beta \equiv E \ ,\ \ \ \ \ \
(B^{\infty})^{2}_{\ 3} =  \tan\alpha \equiv B  \ ,
\nonumber
\ee
\be
(A^{\infty})^0_{\ 1} =-{1\over g_s}{B\over \sqrt{1+B^2}}{1\over \sqrt{1-E^2}}-\chi_0 E  \ , \ \ \ \
\  
(A^{\infty})^{2}_{\ 3} ={1\over g_s}{1\over \sqrt{1+B^2}}{E\over \sqrt{1-E^2}}- \chi_0
B\ ,
\label{asym}
\ee
\be
e^{2\phi_{\infty}}=g_s^2\ ,\ \ \ \ \chi_{\infty}={1\over g_s}{B\over \sqrt{1+B^2}}{E\over
\sqrt{1-E^2}} +\chi_0\ .
\nonumber
\ee
We see that in the $E\to 1$ limit, with 
$G_s, \theta, \tilde\theta$ and $\chi_0$ fixed (see eqs.~(\ref{Ocoup}), (\ref{uuuu})~),
 $A_{01}^\infty$
and
$A_{23}^\infty $ also remain
finite. 

The effective open string dynamics  is governed by the Born-Infeld action
\cite{{FT},{Tsey}} 
$$
S = -{1\over g_s (2\pi )^3 (\a ')^{2} } \int d^4 x  \sqrt{\det
(g+B+2\pi\a ' F)} 
$$
\be
+\pi i \int \left( \chi\  ({B\over 2\pi \a'}+  F)\wedge ({B \over
2\pi\a'}+ F) 
+ 2 (B+2\pi \a' F)\wedge A_2 \right)\ +\int C_4\ .
\label{BI}
\ee
The term involving $\chi $ gives rise to  a  $\Theta$ term in the low energy
field theory, $ \Theta =\chi _\infty \ .$
The Born-Infeld action (\ref{BI}) is invariant
under 
$SL(2,Z)$ transformations \cite{Green}
% (for this it is useful to note that
%  $A_{\mu\nu}=-2{\delta S\over \delta B_{\mu\nu}}$).
(in the low-energy limit this reduces to the Montonen-Olive
duality of the ${\cal N}=4$ SYM theory). 
 The transformation of the fields can be inferred
from the transformation properties under $SL(2,Z)$ of
the type IIB supergravity fields \cite{Green}.

 Under the $SL(2,Z)$ symmetry of the type IIB superstring
the coupling
$$
\lambda=\Theta + {i\over g_s}\ ,\ \ \ \ \ \Theta\equiv \chi_\infty
\ ,
$$
transforms as 
\be\label{sltau}
\lambda\rightarrow \lambda'={a\lambda+b\over c\lambda+d}\ ,\ \ \ \ ad-bc=1\ ,
\ee
where $a,b,c,d$ form an $SL(2,Z)$ matrix, whereas  $B_{\mu\nu}$ and
$A_{\mu\nu}$ (NSNS and RR) fields form a doublet:
\be
\left(\matrix{ B_{\mu\nu}\cr 
A_{\mu\nu}}\right)\rightarrow
\left(\matrix{ B'_{\mu\nu}\cr
A'_{\mu\nu}}\right)=
\left(\matrix{d   & -c\cr
-b & a }\right)
\left(\matrix{ B_{\mu\nu}\cr
A_{\mu\nu}}\right)\ .
\label{ddw}
\ee
Therefore
\be
g_s\rightarrow g_s'= g_s|c\lambda+d|^2\ .
\label{ggs}
\ee
The Einstein metric
$g^{E}_{\mu\nu}=e^{-\phi/2} g^{\rm str}_{\mu\nu}$ remains invariant,
so  the new string metric at $r=\infty $ is 
$|c\lambda+d|\eta_{\mu\nu}$. 
% The factor $|c\lambda+d|$ can be removed by rescaling the coordinates. 
% This leads to a redefinition of the
% fields,  namely $E$ ($B$), as ${E\ (B)\over |c\lambda+d|}$.
Using eqs.~(\ref{asym}),~(\ref{ddw}), one finds that 
the transformed electric and magnetic fields are  
\be\bea{cc}\label{E'B'}
E\rightarrow E'={1\over |c\lambda+d|}\big[ (d+c\Theta )E  +c {B\over G_s}(1-E^2)\big]\ ,
\\
B\rightarrow B'={1\over |c\lambda+d|}\big[  (d+c\Theta )B- c {E\over G_s}(1+B^2)\big]\ .
\label{xxf}
\eea\ee
Let us now consider the  $E\to 1$ limit
for the $SL(2)$ rotated parameters. 
% More precisely, we want to check to what extent 
% $SL(2,Z)$ is also a symmetry of NCOS.
In this limit the transformation (\ref{xxf}) simplifies, since
$g_s\to \infty \ ,\ \lambda\to \Theta $.
One can distinguish two different cases:

\noindent {\it a)} Irrational $\Theta$:

In this case there are no integers $c$ and $d$ such that $c\chi+d=0$.
In the 
$E\to 1$ and $g_s\to \infty$ limit, $|c\lambda+d|$ reduces to
$|c\Theta +d|$ and $E'$ and $B'$ are:
\be\bea{cc}
E'={1\over |c\Theta +d|} (d+c\Theta)E =\pm 1\ ,
\\
B'=\pm B-{c(1+B^2) \over G_s |c\Theta +d| }={\rm
finite}\ .
\eea\ee
In the $E\to 1$ limit, the electric and magnetic fields are obtained by the simple transformation
rules  
$$
{1-{E}^2 \over |c\lambda +d |^2}
\to {1-{E'}^2 \over |c\Lambda+d|^2}\ , \ \ \ \ 
{1+{B}^2 \over |c\lambda +d|^2 }
\to {1+{B'}^2 \over |c\Lambda+d|^2}\ ,
$$ 
where
\be
\Lambda =\chi_0+ {i\over G_s}\ , \ \ \ \ \ \chi_0=\Theta -{B\over G_s}\ .
\ee
The fact that  ${E'}^2\to 1$, with $B'$ finite (and also 
$g'_s\sqrt{1-{E'}^2}$ remains finite) shows that 
an $SL(2,Z)$ transformation leads to another NCOS with 
transformed parameters
$$
(\theta,\td ,G_s,\Theta )\ \longrightarrow \ (\theta ',\td  ',G_s',\Theta ')\ ,
$$ 
where
\be
\Theta '= {a\Theta+b\over c\Theta+d}\ , \ \ \ \ \ 
G_s'=G_s |c \Lambda +d|^2\ .\ \ \
\ee
To find ${\theta'}^{\mu\nu}$ and $G'_{\mu\nu}$  we use
eqs.(\ref{ggg}),(\ref{thth}).
One can choose coordinates (by scaling by suitable factors $\zeta$ and $\eta$ given by
(\ref{zeta})
as in the previous section) so that the open
string metric before the $SL(2)$ transformation is $G_{\mu\nu}=\eta_{\mu\nu} $. 
We find
\be
G'_{\mu\nu}= {|c\Lambda+d|^{2}\over |c\lambda+d|}\ \eta_{\mu\nu}\ ,
\ee
\be
\theta '=2\pi\a ' {E'(1-E^2)\over
|c\lambda+d| (1-{E'}^2)}
= 2\pi\a ' {|c\lambda+d|\over |c\Lambda+d|^2}\ ,
\ee
\be
\tilde \theta '=
2\pi\a ' {B'(1+B^2)\over |c\lambda+d| (1+{B'}^2)}=
2\pi\a 'B' {|c\lambda+d|\over |c\Lambda+d|^2}\ .
\ee
Rescaling the coordinates so that the new open string metric $G'_{\mu\nu}$
is equal to
$\eta_{\mu\nu}$, we conclude that the new NCOS theory has parameters,  
\be
\theta '=2\pi\a '\ , \ \ \ \ \tilde \theta '=2\pi\a '\ B'\ .  
\ee

\noindent {\it b)} Rational $\Theta$:

In this case there exists an $SL(2,Z)$ transformation under which
$c\Theta +d=0$.
The string coupling transforms as follows:
\be
g_s\rightarrow g'_s=g_s|c\lambda+d|^2={c^2\over g_s}\ ,
\ee
i.e. this transformation relates strong and weak coupling regimes.
{} From (\ref{E'B'}) one can find transformed $E$ and $B$ in the $E\to 1$
limit:
\be\bea{cc}
E' = \pm {B\sqrt{1-E^2}\over \sqrt{1+B^2}}
 \to 0\ , \\
B'=\pm {\sqrt{1+B^2}\over \sqrt{1-E^2}}\rightarrow \pm \infty\ .
\eea\ee
Therefore 
\be\bea{cc}
\theta'=2\pi\alpha' {E'(1-E^2)\over |c\lambda+d| (1-{E'}^2)}=0\ , \\
\tilde \theta '=
2\pi\a ' {B'(1+B^2)\over |c\lambda+d| (1+{B'}^2)}=2\pi\alpha' {G_s\over c}={\rm finite},
\eea\ee
and the open string coupling is
\be
G'_s=g'_s\sqrt{1-{E'}^2}\sqrt{1+{B'}^2}=g'_s B'={c^2(1+B^2)\over G_s}={\rm finite}.
\ee

In conclusion, for rational $\Theta$ there is an $SL(2,Z)$ transformation
which maps
the  NCOS theory to  a NCSYM field theory, 
with   $\Theta={a\over c}$. 
In particular, we see that whenever $\Theta=0 $
the NCOS theory is
S-dual to NCSYM theory,
even if both $E$ and $B$ are non-vanishing. 
The reason is that in this case $A_{01}$ given in
(\ref{asym})
also vanishes, and therefore the S-dual theory will have a vanishing $B_{01}$.
This generalizes the result of \cite{GMMS} that the theory with $E=0,\ B\to\infty $ is S-dual to
the theory with $E=1,\ B= 0$.
For irrational $\Theta$, 
under $SL(2,Z)$ transformations 
the NCOS theory is always transformed to a  NCOS theory with
new parameters given
by the above transformation rules.
% eq.~(\ref{qqq}).

%%%%%%%%%%%%%%%%%%%%%%%%%%%%%%%%%%%%%%%%%%%%%%%%%%%%%%%%%%%%%
\newsection{Six-dimensional origin}
%%%%%%%%%%%%%%%%%%%%%%%%%%%%%%%%%%%%%%%%%%%%%%%%%%%%%%%%%%%%

To clarify how these more general NCOS theories fit in the OM theory
framework of \cite{OM}, and
also to exhibit the geometric origin of  $SL(2,Z)$ symmetry,
it is useful to reproduce these theories by starting with  OM theory.
We shall first review  OM-theory and its compactification on an electric
circle   \cite{OM,BBSS}, and 
then  study new compactifications on arbitrary directions of a 2-plane
that lead to the general electric-magnetic NCOS theories 
obtained in sect. 2.

%%%%%%%%%%%%%%%%%%%%%%%%%%%%%%%%%%%%%%%%%%%%%%%%%%%%%%%%%%%%%
\subsection{OM theory on an electric circle}
%%%%%%%%%%%%%%%%%%%%%%%%%%%%%%%%%%%%%%%%%%%%%%%%%%%%%%%%%%%%

We start with a configuration of M5 branes in the presence of a 3-form field 
strength.
For later convenience, we choose the world-volume metric
as follows:
\be
g_{\mu\nu}= {\rm diag}(-\xi^2  ,\xi^2 ,\rho^2 ,\rho^2  ,\rho^2,\xi^2  )\ .
\ee
The field strength is given by
\be
H_{015}=-\xi ^{3} \tanh\beta \ ,\ \ \ \ \ \ \
H_{234}=  \rho ^{3}\sinh\beta\ .
\ee
The $H$ field satisfies the non-linear self-duality constraints \cite{HSW}.
Note that $H$ is rescaled by a factor of $M_P^3$ with respect to
that of ref.~\cite{OM}, so $H$ is dimensionless
($M_P$ is the eleven-dimensional gravitational scale).
Now we take the limit $\beta\to \infty $ with
\be
M_{\rm eff}^3\equiv {1\over 2} M_P ^3\bigg(\xi ^{3}+ H_{015}\bigg)=
{1\over 2} M_P^3 \xi^3 \bigg(1-  \tanh\beta \bigg)  ={\rm fixed}\ ,
\ee
\be
M_P^3\  \xi\rho^2 ={\rm fixed}\ .
\ee
The above quantities represent the effective tension 
(or energy per unit area) of an M2-brane
lying on the 015 and 023 directions, respectively.
This is achieved with
\be
\xi=\xi_0 e^{2\beta/3} \ ,\ \ \ \ \  \rho =\rho_0 e^{-\beta/3}.
\ee
with fixed $\xi_0,\ \rho_0$ and $M_P$ (alternatively, one can send $M_P\to
\infty $ and set $\xi=1 $, as done in \cite{OM}).
As a result, $H_{234}$ remains finite, and
$$
M_{\rm eff}=  M_P \xi_0\ .
$$

Now let us consider  the compactification of OM theory on a circle
parametrized by the $x^5$ direction, $x^5 \equiv x^5+2\pi R_5$.
One obtains  $N$ type IIA D4-branes with a non-zero electric
field along $01$ directions, with the   metric in the string frame given
by
\be
g_{\mu\nu}^{str}=g_s^{2/3}{\rm diag}(-\xi^2,\xi^2,\rho^2,\rho^2,\rho^2)\ ,
\ee
and
\be
g_s=(\xi R_5 M_P)^{3/2} =e^{\beta } (R_5 M_{\rm eff})^{3/2} \ ,\ 
\ \ \ \ \ 
\alpha'= {1\over M_{\rm eff}^3 R_5}\ ,
\label{ppp}
\ee
\be
B_{01}=M_PR_5 \ H_{015}= - M_PR_5\xi^3\tanh\beta\ .
\ee
Thus
\be
g_{\mu\nu}^{str}= \xi_0 R_5 M_P \ {\rm diag}
(-\xi^2_0 e^{2\beta },\xi^2_0 e^{2\beta},\rho^2_0,\rho^2_0,\rho^2_0)\ ,
\ee
\be
E\equiv B^0_{\ 1}= \tanh\beta\ .
\ee
By setting
$$
\xi_0=(4M_PR_5)^{-1/3}\ ,\ \ \ \ \rho_0= 2\xi_0\ ,
$$
the closed string metric becomes 
$g_{\mu\nu}^{str}={\rm diag}(-{1\over 4} e^{2\beta}, {1\over 4}
e^{2\beta},1,1,1)$, and the open string metric is
$$
G_{\mu\nu}= g_{\mu\nu}^{str}- (B g_{\rm str}^{-1}B)_{\mu\nu}=\eta_{\mu\nu}\ .
$$
Therefore the open string coupling $G_s$ is given by
\be
G_s=
g_s \left({\det G_{\mu\nu}\over \det (g_{\mu\nu}+B_{\mu\nu })}
\right)^{1\over 2} = 
g_s\sqrt{1-E^2}=2(M_{\rm eff} R_5)^{3/2}={\rm finite}\ .
\ee
Thus one obtains NCOS theory in $4+1$ dimensions.
Compactifying along the $x^4$ direction
one gets the $3+1$ dimensional NCOS theory with $B_{23}=0$
(which is equivalent to compactifying OM theory on a 2-torus
\cite{OM,BBSS}).

%%%%%%%%%%%%%%%%%%%%%%%%%%%%%%%%%%%%%%%%%%%%%%%%%%%%%%%%%%%%%
\subsection{Oblique reduction of  OM theory}
%%%%%%%%%%%%%%%%%%%%%%%%%%%%%%%%%%%%%%%%%%%%%%%%%%%%%%%%%%%%

Let the coordinates $x_4$ and $x_5$ be periodic with periods
\be
x_5\equiv x_5+2\pi R_5 \ , \ \ \ \ \ \ x_4\equiv x_4+2\pi R_4^0\ .
\ee
Non-vanishing electric and magnetic
fields can be generated by compactifying the theory in an oblique
direction.
Our starting point is the  M5-brane configuration of section
4.1, but with a rotation in the 2-plane $x_4$-$x_5$, i.e.
\be
H_3 =H_{01\tilde 5}\ dx_0\wedge dx_1\wedge d\tilde x_5 +
H_{23\tilde 4}\ dx_2\wedge dx_3\wedge d\tilde x_4\ ,
\label{hhtt}
\ee
$$
\tilde x_4=x_4 \cos\alpha + x_5 \sin\alpha\ ,\ \ \ 
\tilde x_5=- x_4 \sin\alpha +x_5 \cos\alpha\ .
$$
The metric is taken to be
$$
g_{\mu\nu}={\rm diag}(-\xi^2,\xi^2,\rho^2,\rho^2,\xi^2,\xi^2 )\ .
$$
The $H$ field components are
\be
H_{015}=- \xi^3 \tanh\beta \cos\alpha \ ,
\ \ \ \ \ 
H_{014}=\xi^3 \tanh\beta \sin\alpha\ ,
\ee
\be
H_{234}=  \rho^2\xi \sinh\beta \cos\alpha \ ,
\ \ \ \ \
H_{235}=  \rho^2 \xi  \sinh\beta \sin\alpha\ .
\ee

The theory should be decoupled from gravity in  the limit $\beta\to \infty
$ as above, with 
\be\bea{cc}
M_{\rm eff}^3={1\over 2} M_P ^3\bigg(\xi ^{3}+  H_{01\tilde 5}\bigg)=
{1\over 2} M_P^3 \xi^3 \bigg(1-  \tanh\beta \bigg)  ={\rm
fixed}\ , \\
M_P^3 \ \xi\rho^2 ={\rm fixed}\ \ 
\eea\ee
(equivalent conditions can be written, e.g. in terms of $H_{015}$). 
In addition, we also scale $\sin\alpha$ so that
\be
\sinh\beta\ \sin\alpha=B={\rm fixed}\ .
\ee
The above conditions are satisfied by setting
\be
\xi=\xi_0 e^{2\beta/3} \ ,\ \ \ \  \rho =\rho_0 e^{-\beta/3}\ ,\
\ee
with fixed $\xi_0,\ \rho_0,\ B$ and $M_P$.

Dimensional reduction along $x_5$ gives as before $N$ D4 branes with
a $B_{01}$ field, but now there is in addition  a non-vanishing
$B_{23}$ component. Explicitly,
\be\bea{cc}
B_{01}=M_PR_5 \ H_{015}=- M_PR_5\xi^3\tanh\beta\cos\a \ ,\\
B_{23}=M_PR_5\ H_{235} =M_PR_5 \rho^2 \xi\sinh\beta\sin\alpha\ .
\eea\ee
The other components $H_{014},\ H_{234}$ lead to non-zero components
of the 3-form RR field $A_3$. 
The metric is given by
\be
g_{\mu\nu}^{str}=g_s^{2/3}{\rm diag}(-\xi^2,\xi^2,\rho^2,\rho^2,\xi^2)\ ,
\ee
with $g_s$ and $\a' $ given by eq.~(\ref{ppp}).
% \be
% g_s=(\xi R_5 M_P)^{3/2} =e^{\beta } (R_5 M_{\rm eff})^{3/2} \ ,\ 
% \ \ \ \ \ 
% \alpha'= {1\over M_{\rm eff}^3 R_5}\ ,
% \ee
The electric and magnetic field components are given by
\be
B^0_{\ 1}=\tanh\beta\cos\a \ , \ \ \ \ \ B^2_{\ 3}= \sinh\beta\
\sin\alpha\ .
\ee
In the limit $\beta\to\infty $ we get $B^0_{\ 1}=1$ and a magnetic field $B^2_{\ 3}$
equal to $B$. 
Let us write $x_4=R_4^0\theta_4$, where $ \theta_4$ is $2\pi$ periodic. 
In order to have a finite $g_{44}$ component in the $\beta\to \infty $ limit, 
we set
$$
R_4=  {R_4^0\over \sqrt{1-{(B^0_{\ 1})}^2}}={\rm fixed}\ .
$$
% Then $g_{44}=(R_4^0)^2$.
By T-duality in the $x_4$ direction, one obtains $N$ D3 branes,
with non-vanishing NSNS and RR two-form components.  
After setting
$$
\xi_0=[4M_PR_5 (1+B^2)]^{-1/3}\ ,\ \ \ \ \ \rho_0={2\xi_0}
\ ,
$$
the closed string metric for the $3+1$ dimensional world-volume is
$$
g_{\mu\nu}={\rm diag}( -{1\over 1-E^2}\ ,{1\over 1-E^2}\ ,{1\over
1+B^2}\ ,{1\over 1+B^2})\ ,
$$
and hence the open string metric and coupling become
\be
G_{\mu\nu}={\rm diag}(-1,1,1,1)\ , 
\ee
\be
G_s=g^{B}_s\sqrt{1-(B^0_{\ 1})^2}\sqrt{1+(B^2_{\ 3})^2}=
2{R_5\over R_4}(1+B^2)=
{\rm finite}\ .
\ee
where we have used 
$$
g_{44}={R_4^2\over\alpha'}\  ,\ \ \ \ g^B_s={g_s \over \sqrt{g_{44}}}\ .
$$

In the $\beta\to \infty $ limit, the type IIB gauge fields are  as follows:
$$
B^0_{\ 1}=1\ ,\ \ \ \ \ B^2_{\ 3}= B\ ,\ \ \ \ \ \chi=0\ ,
$$
\be
A^0_{\ 1}=0\ ,\ \ \ \ A^2_{\ 3}= {1\over G_s} (1+B^2)\ .
\ee
This agrees with the asymptotic values of the gauge fields for the
corresponding supergravity configuration (\ref{asym}) with $\chi_\infty
=0\ (\chi_0= -{B\over G_s})$.
In order to obtain the more general configuration with non-vanishing
$\chi $, one has to consider the reduction on a non-orthogonal
torus $x_4$-$x_5$, characterized by a modular parameter
$\tau=\tau_1+i\tau_2$.
[Equivalently, one can start with eq.~(\ref{hhtt})
and introduce  coordinates $(\tilde x_4,\tilde x_5)$ by a general
non-orthogonal linear transformation in the $(x_4,x_5)$ plane.]
Then the  metric has a non-vanishing non-diagonal component $g_{45}$ which,
upon dimensional reduction in $x_5$ and T-duality in $x_4$ gives rise to
a non-zero $\chi $ field and also to a non-zero $A_{01}$, in accordance
to eq.~(\ref{asym}).

%%%%%%%%%%%%%%%%%%%%%%%%%%%%%%%%%%%%%%%%%%%%%%%%%%%%%
\newsection{Supergravity description of Large $N$ NCOS}
%%%%%%%%%%%%%%%%%%%%%%%%%%%%%%%%%%%%%%%%%%%%%%%%%%%%%%%%%

The strong coupling $G_s\to\infty $ limit of $U(N)$ NCSYM theory at fixed
$N$
leads to a NCOS theory with $B_{23}=0$. 
Another interesting limit is the  limit of large $N$ with
fixed 't Hooft coupling $2\pi G_sN=g_{YM}^2N$.
For large 't Hooft coupling, this regime can
be studied by using a supergravity description, generalizing
the dualities of \cite{malda}.
The relevant supergravity background was described in \cite{MR}, and it is
obtained from  (\ref{susu}) by scaling the parameters as follows:
\ba
&\cosh \beta  = {  \tilde b' \over \alpha'}\ ,\ \ \ \ \ \
\cos\alpha = {\rm fixed}\ ,
\cr
& x_{0, 1} =  {\tilde b' \over \sqrt{ \alpha'}} \ \tilde x_{0,1} ~,~~~~
x_{2,3} = \sqrt{\alpha' } \cos\alpha \ \tilde x_{2,3} ~, \cr
& r = \sqrt{\alpha'}\  u ~,~~~~~
g = {\tilde b' \cos\alpha\over \alpha' } \hat g\
\cr
 & R^4  = {\rm fixed} = 4 \pi N \hat g\ ,\ \ \ \ \chi_0={\rm fixed}\ .
\ea
One obtains the following supergravity background \cite{MR}: 
$$
ds^2 =  \alpha ' f^{1/2} \bigg[ { u^4 \over R^4}(- d{\tilde x_0}^2 +
 d{\tilde x_1}^2) + { u^4 \over R^4}\hat h (
d{\tilde x_2}^2 +
 d{\tilde x_3}^2) + du^2 + u^2 d\Omega_5^2 \bigg]\ ,
$$
$$
f  = 1 + {R^4 \over u^4 } ~, ~~~~~~~~~
{\hat h}^{-1} = 1 + { u^4 \over R^4 \cos^2\alpha } \ ,
$$
$$
B_{01} = - \alpha' { u^4 \over R^4}  \ ,\ \ \ \ \ \
B_{23} =   \alpha' \tan\alpha { u^4 \over R^4} \hat h \ ,
$$
\be
e^{2\phi} = {\hat g}^2 f^2 {u^8 \over R^8} \hat h\ ,\ \ \ \ \
\chi = { 1 \over \hat g } \tan \alpha f^{-1}+\chi_0\ ,
\label{bbb}
\ee
$$
A_{01} = \alpha'  {u^4 \over R^4}\bigg({1 \over \hat g} \tan\alpha
+\chi_0\bigg)\ , \ \ \ \
\
A_{23} =  \alpha' {u^4 \over R^4}{\hat h} \bigg( {1 \over \hat g} -\chi_0\tan\alpha\bigg) \ ,
$$
$$
F_{0123u}  =  {\alpha'}^2 {1 \over \hat g} \hat h { 4 u^3 \over R^4 } \ ,
\ \ \ \ \ \ R^4=4\pi N\hat g\ .
$$
The coupling constant $\hat g$ can be identified with $G_s$.
It is worth
noting that  
$\tilde x^i,u, R$  are dimensionless. The only 
dimensionful parameter is $\alpha'$, which
is fixed. 

The case $\chi_0 =-{1 \over \hat g} \tan\alpha $ is special.
In this case the $u=\infty $ value of $\chi $ is zero, and
one has $A_{01}=0$.
As a result, the S-dual background (in which $B_{01}$ and $A_{01}$ are exchanged)
has $B_{01}=0$ and describes the purely magnetic case.
This is in agreement with the observation of sect. 3 that the case $\Theta
=0$ is S-dual to NCSYM theory.

%%%%%%%%%%%%%%%%%%%%%%%%%%%%%%%%%%%%%%%%%%%%%%%
\subsection{ Calculation of Wilson loops}
%%%%%%%%%%%%%%%%%%%%%%%%%%%%%%%%%%%%%%%%%%%%%%%

A ``quark-antiquark" potential $V(L)$ for the large $N$ ${\cal N}=4$ SYM
theory 
was computed in \cite{Wilson}, with the result
\be
V(L)=- {c_0\sqrt{g^2_{YM} N }\over L}\ ,\ \ \ \ 
g^2_{YM} N=2\pi \hat g N={1\over 2} R^4\ ,\ \
c_0={4\pi ^2\sqrt{2}\over \Gamma({1\over 4})^4 }\ .
\label{arr}
\ee
The $1/L$ behavior is dictated by the conformal symmetry of the theory.
The fact that $V(L)$ is proportional to   
$\sqrt{ g^2_{YM} N }$ is believed to
be a consequence of strong coupling physics.
Noncommutative super Yang-Mills theory is not scale invariant, 
so the quark-antiquark potential $V(L)$ can have  a more complicated
dependence on $L$ at distances which are comparable with the scale
of noncommutativity. At large distances one should recover the
potential (\ref{arr}) with the usual $-1/L$ behavior.
A question of interest is how the noncommutativity of spacetime
affects the forces between quarks and antiquarks at short distances.
Are these forces weaker or stronger than the commutative case?
At large $N$, this question can be addressed by using
the supergravity description in terms of the background (\ref{bbb}),
and computing a Wilson line.

The case with $B_{01}=0$ was already considered in \cite{MR},
and it was found that strings cannot be localized near the boundary.
In the case of $B_{01}\neq 0$, we will find that if the quark-antiquark
pair lies on a plane orthogonal to the electric field direction,
then the corresponding string
configuration can indeed get to the boundary, and in this way
gives rise to  a quark-antiquark potential.
% By ``generic string" here we mean a quark-antiquark pair which is
% not aligned along the electric field direction.
For pairs which are aligned with the electric field (e.g. 
string configurations of the form $x_0=\tau, x_1=\sigma $) 
the string will stretch from the horizon to the boundary
but will not connect, and hence will not give rise to a potential.
This is expected:  the
electric field tends to separate the charges and for a near critical
electric field it takes 
a lot of energy for the strings to bend and connect.

To illustrate how noncommutative effects modify the
Coulomb potential at short distances, we  consider
 a string configuration in this geometry of the form
$x^0=\tau $, $x^2=\sigma $, ${u\over R}\equiv U=U(\sigma )$, $x_1=x_3=0$
(representing a string  lying on a plane which is orthogonal to
the electric field).
The Nambu-Goto action for this string in the background (\ref{bbb}) is
\be 
S={T\over 2\pi }\int  d\s \sqrt{(1+U^4)\bigg( U^4\hat
h+R^2 (\partial_\s U)^2\bigg)}\ ,\ \ \ \ \hat h={1\over 1+a^4 U^4}\ ,\
\ a^4={1\over \cos ^2\alpha } \ ,
\ee
where the factor $T$ arises from the integral over
the time $x_0$. 
%\be
%E={1\over 2\pi }\int dx \sqrt{f{u^4\over R^4}\bigg({u^4\over R^4}{\hat
%h}+(\partial_x
%u)^2\bigg)}.
%\ee

The solution to the equations of motion for $U(\s )$  is given by
\be
\frac{\sqrt{(1+ U^4)}\ U^4\hat h}
{\sqrt{U^4\hat h+R^2(\partial_{\sigma} U)^2} }  =K={\rm const.}
\ee
Solving the above equation for $\partial_\s U$ we find
\be
R^2(\partial_\s U)^2={U^4\over   K^2 (1+a^4 U^4)^2}\ 
\big[U^8+U^4 - K^2 a^4 U^4 - K^2\big]\ .
\ee
As a function of $\s $, $U$ will vary from $U= \infty $ up to a minimum 
value $U_0$,
given by the solution of $\partial_\s U =0 $.
In terms of $U_0$,  the integration constant $K$ is thus given by
\be
K^2={U^4_0(1+U^4_0)\over 1+a^4 U^4_0}\ .\ \ 
\ee
The integration constant $K$ (or, equivalently, $U_0$) can be determined
by the condition that the
quark
separation is given by some distance $L$,
\be
L(U_0)=\int dx_2= {2 R b\over U_0}
\int_1^{\infty}{dy \over y^2}{1+a^4U_0^4 y^4\over
\sqrt{(y^4-1)(U_0^4 y^4+ b^2)}}\ ,
\label{lle}
\ee
$$
b=\sqrt{ {1+U_0^4\over 1+a^4 U_0^4}}\ .
$$
The  quark-antiquark interaction energy ${\cal E}$ then follows
from the relation $S=T{\cal E}$. In this expression, one should subtract
as in \cite{Wilson} the ``quark masses" corresponding to the energy of two
strings that stretch from the horizon to infinity. 
We obtain
$$
V(U_0)={\cal E}(U_0)-{\cal E}(0)={RU_0 \over \pi}
\int_1^{\infty} dy
\sqrt{1+U_0^4y^4}\left( 
{y^2 \sqrt{1+U_0^4y^4}\over \sqrt{(y^4-1)(U_0^4 y^4+ b^2)}} -1\right)
$$
\be
- {RU_0 \over \pi}   \int_0^1 dy \sqrt{1+U_0^4y^4}\ .
\label{enn}
\ee
Equations (\ref{lle}), ( \ref{enn}) define the quark-antiquark
potential $V(L)$.

Let us first examine $V(L)$ in the case of large distances $L\gg 1$.
Restoring $\a '$, this means $L\gg \sqrt{\a' }$, or $L^2\gg \theta $, 
where $\theta $ is the noncommutative scale parameter in the $x_0$-$x_1$
plane,
given by $\theta=2\pi\a '$.
As a function of $U_0$, $L(U_0)$ 
has a minimum value
$L_{\rm min}$ at some $U_0=U_1$, and it goes to infinity in both the small
and large $U_0$ regions 
(for small $U_0$ one has $L\sim 1/U_0$;  for large $U_0$ one has
$L\sim U_0$). Thus there are two possible values of $U_0$ for
a given $L>L_{\rm min}$, and hence two possible values for the potential
for a given
$L$.  It is easy to see that the branch $U_1<U_0<\infty $ has higher
energy and it is
unstable, so we restrict to the branch $0<U_0<U_1$.
The large distance region is then obtained in the region of small $U_0$.
To obtain the potential $V$  as a function of $L$, we need to  find
$U_0=U_0(L)$ using  (\ref{lle}). Although the integral (\ref{lle}) cannot
be performed analytically, a systematic (large distance) expansion in
powers of $1/L$ can be obtained by expanding
the factor $1/\sqrt{y^4-1} $ in powers of $1/y$.
Consider for example the purely electric case corresponding
to $\alpha=0$  (and hence $a=b=1$).
The resulting integrals are hypergeometric functions which can
be expanded in powers of $U_0$. A given order in $U_0$ is given by
a series which can be re-summed.
We obtain the following result:
\be
{L\over R}={ a_0\over U_0} + a_1 U_0^2 +O(U_0^3)\ ,
\ee
$$
a_0=2\int_1^\infty {dy\over y^2}{1\over \sqrt{y^4-1}
}={2\sqrt{2}\pi^{3/2}\over
\Gamma({1\over 4})^2 }\ ,\ \ \ \ \ a_1={\Gamma({1\over 4})^2\over
3\sqrt{\pi } }\ ,
$$
or
\be
U_0={a_0R\over L}+ a_1 a_0^3 {R^4\over L^4} +O(1/L^5)\ .
\label{zxx}
\ee
Expanding the integral eq.~(\ref{enn}) in powers of $U_0$ and inserting
eq.~(\ref{zxx})  we obtain
\be
V(L)=- {c_0\sqrt{g^2_{YM}N } \over L} -  c_1{(2g^2_{YM}N)^{5/4}\over L^4}
+O(1/L^5)\ ,
\label{vvll}
\ee
$$
c_0={4\pi ^2\sqrt{2}\over \Gamma({1\over 4})^4 }\ ,\ \ \ \ 
c_1={16 \pi^{9/2}\over 3 \Gamma({1\over 4})^6 }\ .
$$
{}From this formula we can read the first corrections
to a quark-antiquark potential due to the noncommutative nature of 
spacetime. 
Comparing this potential with the usual SYM case (represented by 
the first
term in (\ref{vvll})~),
we see that the force between
oppositely
charged particles increases more rapidly
as the distance is decreased from $L\gg \sqrt{\a' }$.

\begin{figure}
\PSbox{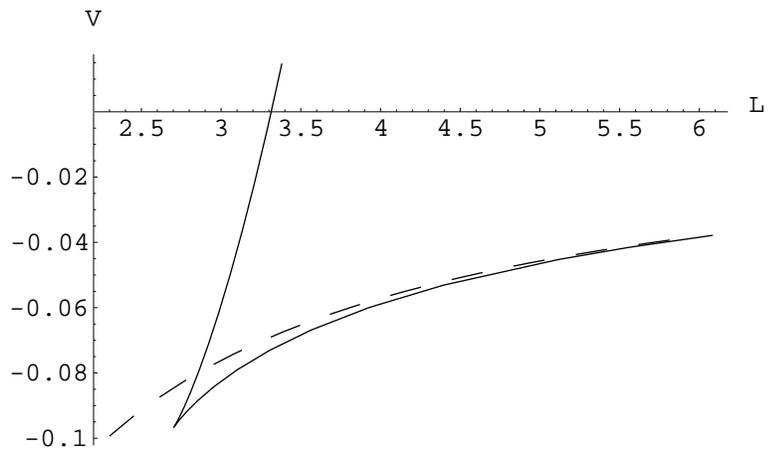 
% hscale=70 vscale=70 
% hoffset=50 voffset=-60
} {15cm}{8cm}
\caption{
The quark-antiquark potential as a function of the distance $L$.
The numerical plot is done for $R=1$, $\alpha'=1 $, $a=1$.
The solid line crossing $V=0$ is unphysical, describing the unstable branch.
The dashed line represents the Coulombian potential $-c_0/L$ of the
usual ${\cal N}=4$ SYM theory.
\label{fig:wilson}}
\end{figure}

At shorter distances, the potential $V(L)$ is more complicated, 
but it can be computed numerically.
Figure 1 shows a plot of $V(L)$ for the ordinary SYM theory and 
for the present noncommutative case.
The curve $V(L)$ terminates at $L=L_{\rm min}=L(U_1)$ ($\cong 2.7$ in the figure).
For shorter distances, there is no solution.
 Presumably, this may indicate a breakdown of
   the present
simple description at very short distances.
{}From the figure, we can see that  the net effect of   
noncommutativity is increasing the force between a quark and an
antiquark.
A similar behavior is obtained for a generic  $\alpha\neq \pi/2$, 
i.e. configurations with both electric and magnetic fields.

\bigskip\bigskip\bigskip

{\bf Acknowledgements}

 J.R. would like to thank  ICTP for hospitality, where this work
was carried out.
The work of M.M. Sh-J. was partly supported by the EC contract
no. ERBFMRX-CT 96-0090.

\end{document}